\documentclass[conference]{IEEEtran}
\makeatletter
\def\ps@headings{%
\def\@oddhead{\mbox{}\scriptsize\rightmark \hfil \thepage}%
\def\@evenhead{\scriptsize\thepage \hfil \leftmark\mbox{}}%
\def\@oddfoot{}%
\def\@evenfoot{}}
\makeatother 

\usepackage{graphicx}
\usepackage{subfigure}
\usepackage{color}
\usepackage{epsfig}
\usepackage{amssymb}
\usepackage{latexsym,url}

\newcommand{\probconverge}{\mbox{~$\stackrel{P}{\rightarrow}$~}}

\newcommand {\C} {{\rm I\kern-5.5pt C}}

\newcommand{\bP}[1]{{\mathbb{P}}\left[{#1}\right]}
\newcommand{\bE}[1]{{\mathbb{E}}\left[{#1}\right]}

\newcommand{\1}[1]{{\bf 1}\left[#1\right]}       

\newcommand{\fsquare}{\vrule height6pt width7pt depth1pt}   
\newcommand{\myendpf}{\hfill\fsquare \\[0.2cm]}             

\newtheorem{theorem}{Theorem}[section]

\newtheorem{lemma}[theorem]{Lemma}

\newtheorem{corollary}[theorem]{Corollary}

\begin{document}

\title{On the gradual deployment of random \\pairwise key distribution schemes}

\author{
\authorblockN{Osman Ya\u{g}an and Armand M. Makowski}
\authorblockA{Department of Electrical and Computer Engineering\\
              and the Institute for Systems Research\\
              University of Maryland, College Park\\
              College Park, Maryland 20742\\
              oyagan@umd.edu, armand@isr.umd.edu}
}

\maketitle

\begin{abstract}
\normalsize In the context of wireless sensor networks, the
pairwise key distribution scheme of Chan et al. has several
advantages over other key distribution schemes including the
original scheme of Eschenauer and Gligor. However, this offline
pairwise key distribution mechanism requires that the network size
be set in advance, and involves all sensor nodes simultaneously.
Here, we address this issue by describing an implementation of the
pairwise scheme that supports the gradual deployment of sensor
nodes in several consecutive phases. We discuss the key ring size
needed to maintain the secure connectivity throughout all the
deployment phases. In particular we show that the number of keys
at each sensor node can be taken to be $O(\log n)$ in order to
achieve secure connectivity (with high probability).
\end{abstract}


\section{Introduction}
\label{sec:Introduction}

Wireless sensor networks (WSNs) are distributed collections of
sensors with limited capabilities for computations and wireless
communications. Such networks will likely be deployed in hostile
environments where cryptographic protection will be needed to
enable secure communications, sensor-capture detection, key
revocation and sensor disabling. However, traditional key exchange
and distribution protocols based on trusting third parties have
been found inadequate for large-scale WSNs, e.g., see
\cite{EschenauerGligor,PerrigStankovicWagner,SunHe} for
discussions of some of the challenges.

{\em Random} key predistribution schemes were recently proposed to
address some of these challenges. The idea of randomly assigning
secure keys to sensor nodes prior to network deployment was first
proposed by Eschenauer and Gligor \cite{EschenauerGligor}. The
modeling and performance of the EG scheme, as we refer to it
hereafter, has been extensively investigated \cite{BlackburnGerke,
DiPietroManciniMeiPanconesiRadhakrishnan2008, EschenauerGligor,
Rybarczyk2009, YaganMakowskiISIT2008, YaganMakowskiISIT2009,
YaganMakowskiConnectivity}, with most of the focus being on the
{\em full visibility} case where nodes are all within
communication range of each other. Under full visibility, the EG
scheme induces so-called {\em random key graphs}
\cite{YaganMakowskiISIT2008} (also known in the literature as {\em
uniform} random intersection graphs \cite{BlackburnGerke}).
Conditions on the graph parameters to ensure the absence of
isolated nodes have been obtained independently in
\cite{BlackburnGerke,YaganMakowskiISIT2008} while the papers
\cite{BlackburnGerke,
DiPietroManciniMeiPanconesiRadhakrishnan2008, Rybarczyk2009,
YaganMakowskiISIT2009, YaganMakowskiConnectivity} are concerned
with zero-one laws for connectivity. Although the assumption of
full visibility does away with the wireless nature of the
communication infrastructure supporting WSNs, in return this
simplification makes it possible to focus on how randomizing the
key selections affects the establishment of a secure network; the
connectivity results for the underlying random key graph then
provide helpful (though optimistic) guidelines to dimension the EG
scheme.

The work of Eschenauer and Gligor has spurred the development of
other key distribution schemes which perform better than the EG
scheme in some aspects, e.g., \cite{ChanPerrigSong,
DuDengHanVarshney, PerrigStankovicWagner,SunHe}. Although these
schemes somewhat improve resiliency, they fail to provide {\em
perfect} resiliency against node capture attacks. More
importantly, they do not provide a node with the ability to
authenticate the identity of the neighbors with which it
communicates. This is a major drawback in terms of network
security since {\em node-to-node authentication} may help detect
node misbehavior, and provides resistance against node replication
attacks \cite{ChanPerrigSong}.

To address this issue Chan et al. \cite{ChanPerrigSong} have
proposed the following  random {\em pairwise} key predistribution
scheme: Before deployment, each of the $n$ sensor nodes is paired
(offline) with $K$ distinct nodes which are randomly selected
amongst all other $n-1$ nodes. For each such pairing, a unique
pairwise key is generated and stored in the memory modules of each
of the paired sensors along with the id of the other node. A
secure link can then be established between two communicating
nodes if at least one of them has been assigned to the other,
i.e., if they have at least one key in common. See Section
\ref{sec:ModelPairwise} for implementation details.

This scheme has the following advantages over the EG scheme (and
others): (i) Even if some nodes are captured, the secrecy of the
remaining nodes is {\em perfectly} preserved; and (ii) Unlike
earlier schemes, this scheme enables both node-to-node
authentication and quorum-based revocation without involving a
base station. Given these advantages, we found it of interest to
model the pairwise scheme of Chan et al. and to assess its
performance. In \cite{YaganMakowskiRandomPairwise} we began a
formal investigation along these lines. Let $\mathbb{H}(n;K)$
denote the random graph on the vertex set $\{ 1, \ldots , n \}$
where distinct nodes $i$ and $j$ are adjacent if they have a
pairwise key in common; as in earlier work on the EG scheme this
corresponds to modeling the random pairwise distribution scheme
under full visibility. In \cite{YaganMakowskiRandomPairwise} we
showed that the probability of $\mathbb{H}(n;K)$ being connected
approaches $1$ (resp. $0$) as $n$ grows large if $K \geq 2$ (resp.
if $K=1$), i.e., $\mathbb{H}(n;K)$ is asymptotically almost surely
(a.a.s.) connected whenever $K \geq 2$.

In the present paper, we continue our study of connectivity
properties but from a different perspective: We note that in many
applications, the sensor nodes are expected to be deployed
gradually over time. Yet, the pairwise key distribution is an {\em
offline} pairing mechanism which simultaneously involves all $n$
nodes. Thus, once the network size $n$ is set, there is no way to
add more nodes to the network and still {\em recursively} expand
the pairwise distribution scheme (as is possible for the EG
scheme). However, as explained in Section
\ref{subsec:MultiplePhasesImplementingPairwise}, the gradual
deployment of a large number of sensor nodes is nevertheless
feasible from a practical viewpoint. In that context we are
interested in understanding how the parameter $K$ needs to scale
with $n$ large in order to ensure that connectivity is {\em
maintained} a.a.s. throughout gradual deployment. We also discuss
the number of keys needed in the memory module of each sensor to
achieve secure connectivity at every step of the gradual
deployment. Since sensor nodes are expected to have very limited
memory, it is crucial for a key distribution scheme to have {\em
low} memory requirements \cite{DuDengHanVarshney}.

The key contributions of the paper can be stated as follows: Let
$\mathbb{H}_{\gamma}(n;K)$ denote the subgraph of
$\mathbb{H}(n;K)$ restricted to the nodes $1, \ldots, \lfloor
\gamma n \rfloor$. We first present scaling laws for the absence
of isolated nodes in the form of a full zero-one law, and use
these results to formulate conditions under which
$\mathbb{H}_{\gamma}(n;K)$ is a.a.s. {\em not} connected. Then,
with $0< \gamma_1 < \gamma_2< \ldots < \gamma_{\ell}<1$, we give
conditions on $n$, $K$ and $\gamma_1$ so that
$\mathbb{H}_{\gamma_i}(n;K)$ is a.a.s. connected for each $i=1,2,
\ldots, \ell$; this corresponds to the case where the network is
connected in {\em each} of the $\ell$ phases of the gradual
deployment. As with the EG scheme, these scaling conditions can be
helpful for dimensioning the pairwise key distribution in the case
of gradual deployment. We also discuss the required number of keys
to be kept in the memory module of each sensor to achieve secure
connectivity at every step of the gradual deployment. Since sensor
nodes are expected to have very limited memory, it is crucial for
a key distribution scheme to have {\em low} memory requirements
\cite{DuDengHanVarshney}. In contrast with the EG scheme (and its
variants), the key rings produced by the pairwise scheme of Chan
et al. have variable size between $K$ and $K+(n-1)$. Still, we
show that the {\em maximum} key ring size is on the order $\log n$
with very high probability provided $K=O(\log n)$. Combining with
the connectivity results, we conclude that the sensor network can
maintain the a.a.s. connectivity through all phases of the
deployment when the number of keys to be stored in each sensor's
memory is $O(\log n)$; this is a key ring size comparable to that
of the EG scheme (in realistic WSN scenarios
\cite{DiPietroManciniMeiPanconesiRadhakrishnan2008}).

These results show that the pairwise scheme can also be feasible
when the network is deployed gradually over time. However, as with
the results in \cite{YaganMakowskiRandomPairwise}, the assumption
of full visibility may yield a dimensioning of the pairwise scheme
which is too optimistic. This is due to the fact that the
unreliable nature of wireless links has not been incorporated in
the model. However the results obtained in this paper already
yield a number of interesting observations: The obtained zero-one
laws differ significantly from the corresponding results in the
single deployment case \cite{YaganMakowskiRandomPairwise}. Thus,
the gradual deployment may have a significant impact on the
dimensioning of the pairwise distribution algorithm. Yet, the
required number of keys to achieve secure connectivity being
$O(\log n)$, it is still feasible to use the pairwise scheme in
the case of gradual deployment; the required key ring size in EG
scheme is also $O(\log n)$ under full-visibility
\cite{DiPietroManciniMeiPanconesiRadhakrishnan2008}.


\section{The model}
\label{sec:ModelPairwise}

\subsection{Implementing pairwise key distribution schemes}
\label{subsec:ImplementingPairwise}

The random pairwise key predistribution scheme of Chan et al. is
parametrized by two positive integers $n$ and $K$ such that $K <
n$. There are $n$ nodes which are labelled $i=1, \ldots , n$. with
unique ids ${\rm Id}_1, \ldots , {\rm Id}_n$. Write ${\cal N} :=
\{ 1, \ldots n \}$ and set ${\cal N}_{-i} := {\cal N}-\{i\}$ for
each $i=1, \ldots , n$. With node $i$ we associate a subset
$\Gamma_{n,i}$ nodes selected at {\em random} from ${\cal N}_{-i}$
-- We say that each of the nodes in $\Gamma_{n,i}$ is paired to
node $i$. Thus, for any subset $A \subseteq {\cal N}_{-i}$, we
require
\[
\bP{ \Gamma_{n,i} = A } = \left \{
\begin{array}{ll}
{{n-1}\choose{K}}^{-1} & \mbox{if $|A|=K$} \\
              &                   \\
0             & \mbox{otherwise.}
\end{array}
\right .
\]
The selection of $\Gamma_{n,i}$ is done {\em uniformly} amongst
all subsets of ${\cal N}_{-i}$ which are of size exactly $K$. The
rvs $\Gamma_{n,1}, \ldots , \Gamma_{n,n}$ are assumed to be
mutually independent so that
\[
\bP{ \Gamma_{n,i} = A_i, \ i=1, \ldots , n } = \prod_{i=1}^n \bP{
\Gamma_{n,i} = A_i }
\]
for arbitrary $A_1, \ldots , A_n$ subsets of ${\cal N}_{-1},
\ldots , {\cal N}_{-n} $, respectively.

On the basis of this {\em offline} random pairing, we now
construct the key rings $\Sigma_{n,1}, \ldots , \Sigma_{n,n}$, one
for each node, as follows: Assumed available is a collection of
$nK$ distinct cryptographic keys $\{ \omega_{i|\ell}, \ i=1,
\ldots , n ; \ \ell=1, \ldots , K \}$ -- These keys are drawn from
a very large pool of keys; in practice the pool size is assumed to
be much larger than $nK$, and can be safely taken to be infinite
for the purpose of our discussion.

Now, fix $i=1, \ldots , n$ and let $\ell_{n,i}: \Gamma_{n,i}
\rightarrow \{ 1, \ldots , K \}$ denote a labeling of
$\Gamma_{n,i}$. For each node $j$ in $\Gamma_{n,i}$ paired to $i$,
the cryptographic key $\omega_{i|\ell_{n,i}(j)}$ is associated
with $j$. For instance, if the random set $\Gamma_{n,i}$ is
realized as $\{ j_1, \ldots , j_K \}$ with $1 \leq j_1 < \ldots <
j_K \leq n $, then an obvious labeling consists in
$\ell_{n,i}(j_k) = k $ for each $k=1, \ldots , K$ with key
$\omega_{i|k}$ associated with node $j_k$. Of course other
labeling are possible.  e.g., according to decreasing labels or
according to a random permutation. The pairwise key
\[
\omega^\star_{n,ij} = [ {\rm Id}_i | {\rm Id}_j |
\omega_{i|\ell_{n,i}(j)} ]
\]
is constructed and inserted in the memory modules of both nodes
$i$ and $j$. Inherent to this construction is the fact that the
key $\omega^\star_{n,ij}$ is assigned {\em exclusively} to the
pair of nodes $i$ and $j$, hence the terminology pairwise
distribution scheme. The key ring $\Sigma_{n,i}$ of node $i$ is
the set
\begin{equation}
\Sigma_{n,i} := \{ \omega^\star_{n,ij}, \ j \in \Gamma_{n,i} \}
\cup \{ \omega^\star_{n,ji}, \ i \in \Gamma_{n,j} \}
\label{eq:KeyRingDefn}
\end{equation}
as we take into account the possibility that node $i$ was paired
to some other node $j$. As mentioned earlier, under full
visibility, two node, say $i$ and $j$, can establish a secure link
if at least one of the events $i \in \Gamma_{n,j}$ or $j \in
\Gamma_{n,j}$ is taking place. Note that both events can take
place, in which case the memory modules of node $i$ and $j$ each
contain the distinct keys $\omega^\star_{n,ij}$ and
$\omega^\star_{n,ji}$. It is also plain that by construction this
scheme supports node-to-node authentication.

\subsection{Gradual deployment}
\label{subsec:MultiplePhasesImplementingPairwise}

Initially $n$ node identities were generated and the key rings
$\Sigma_{n,1}, \ldots, \Sigma_{n,n}$ were constructed as indicated
above -- Here $n$ stands for the maximum possible network size and
should be selected large enough. This key selection procedure does
not require the physical presence of the sensor entities and can
be implemented completely on the software level. We now describe
how this offline pairwise key distribution scheme can support
gradual network deployment in consecutive stages. In the initial
phase of deployment, with $0 < \gamma_1 < 1$, let $\lfloor
\gamma_1 n \rfloor $ sensors be produced and given the labels
$1,\ldots, \lfloor \gamma_1 n \rfloor$. The key rings
$\Sigma_{n,1}, \ldots, \Sigma_{n,\lfloor \gamma_1 n \rfloor}$ are
then inserted into the memory modules of the sensors $1,\ldots,
\lfloor \gamma_1 n \rfloor$, respectively. Imagine now that more
sensors are needed, say $\lfloor \gamma_2 n \rfloor - \lfloor
\gamma_1 n \rfloor $ sensors with $0 < \gamma_1 < \gamma_2 \leq
1$. Then, $\lfloor \gamma_2 n \rfloor - \lfloor \gamma_1 n \rfloor
$ additional sensors would be produced, this second batch of
sensors would be assigned labels $\lfloor \gamma_1 n \rfloor +1,
\ldots , \lfloor \gamma_2 n \rfloor$, and the key rings
$\Sigma_{n,\lfloor \gamma_1 n \rfloor +1}, \ldots ,
\Sigma_{n,\lfloor \gamma_2 n \rfloor}$ would be inserted into
their memory modules. Once this is done, these $\lfloor \gamma_2 n
\rfloor  - \lfloor \gamma_1 n \rfloor $ new sensors are added to
the network (which now comprises $\lfloor \gamma_2 n \rfloor $
deployed sensors). This step may be repeated a number times: In
fact, for some finite integer $\ell$, consider positive scalars $0
< \gamma_1 <  \ldots < \gamma_\ell \leq 1 $ (with $\gamma_0 = 0$
by convention). We can then deploy the sensor network in $\ell$
consecutive phases, with the $k^{th}$ phase adding $\lfloor
\gamma_k n \rfloor - \lfloor \gamma_{k-1} n \rfloor $ new nodes to
the network for each $k=1, \ldots , \ell$.

\section{Related work}
\label{sec:RelatedWork}

The pairwise distribution scheme naturally gives rise to the
following class of random graphs: With $n=2,3, \ldots $ and
positive integer $K$ with $K<n$, the distinct nodes $i$ and $j$
are said to be {\em adjacent}, written $i \sim j$, if and only if
they have at least one key in common in their key rings, namely
\begin{equation}
i \sim j \quad \mbox{iff} \quad \Sigma_{n,i} \cap \Sigma_{n,j}
\neq \emptyset . \label{eq:Adjacency}
\end{equation}
Let $\mathbb{H}(n;K)$ denote the undirected random graph on the
vertex set $\{ 1, \ldots , n \}$ induced through the adjacency
notion (\ref{eq:Adjacency}). With $P (n;K) := \bP{ \mathbb{H}(n;
K) ~\mbox{is connected} }$, we have shown
\cite{YaganMakowskiRandomPairwise} the following zero-one law.

\begin{theorem}
{\sl With $K$ a positive integer, it holds that
\begin{equation}
\lim_{n \rightarrow \infty } P(n;K) = \left \{
\begin{array}{ll}
0 & \mbox{if~ $K = 1$} \\
1 & \mbox{if~$K \geq 2$.}
\end{array}
\right . \label{eq:OneLaw+Connectivity}
\end{equation}
Moreover, for any $K \geq 2$, we have
\begin{equation}
P(n;K) \geq 1 - \frac{27}{2n^2}
\label{eq:OneLaw+BoundForConnectivity}
\end{equation}
for all $n=2,3, \ldots $ sufficiently large.}
\label{thm:OneLaw+Connectivity}
\end{theorem}

\section{The results}
\label{sec:Results}

We now present the main results of the paper. We start with the
results regarding the key ring sizes: Theorem
\ref{thm:OneLaw+Connectivity} shows that very small values of $K$
suffice for a.a.s. connectivity of the random graph
$\mathbb{H}(n;K)$. The mere fact that $\mathbb{H}(n;K)$ becomes
connected even with very small $K$ values does not imply that the
{\em number} of keys (i.e., the size $|\Sigma_{n,i}|$) to achieve
connectivity is necessarily small. This is because in contrast
with the EG scheme and its variants, the pairwise scheme produces
key rings of variable size between $K$ and $K +(n-1)$. To explore
this issue further we first obtain minimal conditions on a scaling
$K: \mathbb{N}_0 \rightarrow \mathbb{N}_0$ which ensure that the
key ring of a node has size roughly of the order (of its mean)
$2K_n$ when $n$ is large.

\begin{lemma}
{\sl For any scaling $K: \mathbb{N}_0 \rightarrow \mathbb{N}_0$,
we have
\begin{equation}
 \frac{|\Sigma_{n,1}(K_n)|}{2K_n} {\probconverge}_n ~1
\label{eq:keyring_lemma}
\end{equation}
as soon as $\lim_{n \rightarrow \infty} K_n = \infty$. }
\label{lem:KeyRing}
\end{lemma}

Thus, when $n$ is large $|\Sigma_{n,1}(K_n)|$ fluctuates from
$K_n$ to $K_n +(n-1)$ with a propensity to hover about $2K_n$
under the conditions of Lemma \ref{lem:KeyRing}. This result is
sharpened with the help of a concentration result for the maximal
key ring size under an appropriate class of scalings. We define
the maximal key ring size by
\[
M_n := \left ( \max_{i=1, \ldots , n }  |\Sigma_{n,i}| \right ),
\quad n=2,3, \ldots
\]

\begin{theorem}
{\sl Consider a scaling $K: \mathbb{N}_0 \rightarrow \mathbb{N}_0$
of the form
\begin{equation}
K_n \sim \lambda  \log n
\label{eq:FormOfScaling}
\end{equation}
with $\lambda> 0$. If $\lambda > \lambda^\star := \left(2 \log 2
-1\right)^{-1}\simeq 2.6$, then there exists $c(\lambda)$ in the
interval $(0,\lambda)$ such that
\begin{equation}
\lim_{n \rightarrow \infty } \bP{ \left | M_n  (K_n) - 2 K_n
\right | \geq c \log n } = 0 \label{eq:KeyRingSize}
\end{equation}
whenever $ c(\lambda) < c < \lambda$. } \label{thm:KeyRingSize}
\end{theorem}
In the course of proving Theorem \ref{thm:KeyRingSize} we also
show that
\begin{equation}
\bP{ \left | M_n  (K_n) - 2 K_n \right | \geq c \log n } \leq
2n^{-h(\gamma;c)} \label{eq:KeyRingSizeBound}
\end{equation}
for all $n=1,2, \ldots$ whenever $ c(\gamma) < c < \gamma$ with
$h(\gamma ; c) > 0$ specified in
\cite{YaganMakowskiRandomPairwise}.

With the network deployed gradually over time as described in
Section \ref{sec:ModelPairwise}, we are now interested in
understanding how the parameter $K$ needs to be scaled with large
$n$ to ensure that connectivity is {\em maintained} a.a.s.
throughout gradual deployment. Consider positive integers $n=2,3,
\ldots $ and $K$ with $K<n$. With $\gamma$ in the interval
$(0,1)$, let $\mathbb{H}_{\gamma}(n;K)$ denote the subgraph of
$\mathbb{H}(n;K)$ restricted to the nodes $\{ 1, \ldots , \lfloor
\gamma n \rfloor \}$. Given scalars $0 < \gamma_1 < \ldots <
\gamma_\ell \leq 1 $, we seek conditions on the parameters $K$ and
$n$ such that $\mathbb{H}_{\gamma_i}(n;K)$ is a.a.s. connected for
each $i=1, 2, \ldots, \ell$.

First we write $P_\gamma (n;K) := \bP{ \mathbb{H}_\gamma(n; K)
~\mbox{is connected} } = \bP{ C_{n,\gamma} (K) }$ with
$C_{n,\gamma} (K)$ denoting the event that $\mathbb{H}_\gamma
(n;K)$ is connected. The fact that $\mathbb{H} (n; K)$ is
connected does {\em not} imply that $\mathbb{H}_\gamma (n; K)$ is
necessarily connected. Indeed, with distinct nodes $i,j=1, \ldots
, \lfloor \gamma n \rfloor $, the path that exists in $\mathbb{H}
(n; K)$ between these nodes (as a result of the assumed
connectivity of $\mathbb{H} (n; K)$) may comprise edges that are
not in $\mathbb{H}_\gamma (n; K)$. The next result provides an
analog of Theorem \ref{thm:OneLaw+Connectivity} in this new
setting.

\begin{theorem}
{\sl With $\gamma$ in the unit interval $(0,1)$ and $c > 0$,
consider a scaling $K: \mathbb{N}_0 \rightarrow \mathbb{N}_0$ such
that
\begin{equation}
K_n \sim c ~ \frac{\log n}{\gamma} . \label{eq:scaling_K}
\end{equation}
Then, we have $ \lim_{n \rightarrow \infty} P_\gamma (n; K_n) = 1
\label{eq:OneLaw_Partial_Deployment} $
 whenever $c >
1$.} \label{thm:OneLaw_Partial_Deployment}
\end{theorem}

The random graphs $\mathbb{H} (n; K)$ and $\mathbb{H}_\gamma (n;
K)$ have very different neighborhood structures. Indeed, any node
in $\mathbb{H} (n; K)$ has degree at least $K$, so that no node is
isolated in $\mathbb{H} (n; K)$. However, there is a positive
probability that isolated nodes exist in $\mathbb{H}_\gamma (n;
K)$. In fact, with $P^{\star}_\gamma (n; K_n) := \bP{
\mathbb{H}_\gamma (n; K) ~\mbox{contains no isolated nodes} }$, we
have the following zero-one law.

\begin{theorem}
{\sl With $\gamma$ in the unit interval $(0,1)$, consider a
scaling $K: \mathbb{N}_0 \rightarrow \mathbb{N}_0$ such that
(\ref{eq:scaling_K}) holds for some $c>0$. Then, we have
\begin{equation}
\lim_{n \rightarrow \infty} P^{\star}_\gamma (n; K_n)
 = \left \{
\begin{array}{ll}
0 & \mbox{if~ $c < r(\gamma) $} \\
1 & \mbox{if~$c>r(\gamma)$}
\end{array}
\right . \label{eq:Isol_Partial_Deployment}
\end{equation}
where the threshold $r(\gamma)$ is given by
\begin{equation}
r(\gamma) := \left ( 1-\frac{\log(1-\gamma)}{\gamma} \right
)^{-1}. \label{eq:r_gamma_defn}
\end{equation}
} \label{thm:Isol_Partial_Deployment}
\end{theorem}

It is easy to check that $r(\gamma)$ is decreasing on the interval
$[0,1]$ with $\lim_{\gamma \downarrow 0}r(\gamma) = \frac{1}{2}$
and $\lim_{\gamma \uparrow 1}r(\gamma) = 0$. Since a connected
graph has no isolated nodes, Theorem
\ref{thm:Isol_Partial_Deployment} yields $\lim_{n \rightarrow
\infty} \bP{ \mathbb{H}_\gamma(n; K_n) ~{\rm is~connected}} =0 $
if the scaling $K: \mathbb{N}_0 \rightarrow \mathbb{N}_0$
satisfies (\ref{eq:scaling_K}) with $c < r(\gamma)$. The following
corollary is now immediate from Theorem
\ref{thm:OneLaw_Partial_Deployment}.

\begin{corollary}
{\sl With $\gamma$ in the unit interval $(0,1)$, consider a
scaling $K: \mathbb{N}_0 \rightarrow \mathbb{N}_0$ such that
(\ref{eq:scaling_K}) holds for some $c>0$. Then, with $r(\gamma)$
given by (\ref{eq:r_gamma_defn}), we have
\begin{equation}
\lim_{n \rightarrow \infty} P_\gamma (n; K_n)
 = \left \{
\begin{array}{ll}
0 & \mbox{if~ $c < r(\gamma) $} \\
1 & \mbox{if~$c>1$}
\end{array}
\right . \label{eq:connec_zero_one}
\end{equation}
 } \label{cor:connectivity}
\end{corollary}

Corollary \ref{cor:connectivity} does not provide a full zero-one
law for the connectivity of $\mathbb{H}_\gamma(n; K_n)$ as there
is a gap between the threshold $r(\gamma)$ of the zero-law and the
threshold $1$ of the one-law. Yet, the gap between the thresholds
of the zero-law and the one-law is quite small with $\frac{1}{2} <
1 - r(\gamma) < 1$. More importantly, Corollary
\ref{cor:connectivity} already implies (via a monotonicity
argument) that it is necessary {\em and} sufficient to keep the
parameter $K_n$ on the order of $\log n$ to ensure that the graph
$\mathbb{H}_\gamma(n; K_n)$ is a.a.s. connected. It is worth
pointing out that the simulation results in Section
\ref{sec:Simulation} suggest the existence of a full zero-one law
for $P_\gamma(n; K_n)$ with a threshold resembling $r(\gamma)$.
This would not be surprising since in many known classes of random
graphs, the absence of isolated nodes and graph connectivity are
asymptotically equivalent properties, e.g., Erd\H{o}s-R\'enyi
graphs \cite{Bollobas} and random key graphs \cite{Rybarczyk2009},
among others.

Finally we turn to gradual network deployment as discussed in
Section \ref{sec:ModelPairwise}.

\begin{theorem}
{\sl With $0 < \gamma_1 < \gamma_2 < \ldots < \gamma_{\ell} \leq 1
$, consider a scaling $K: \mathbb{N}_0 \rightarrow \mathbb{N}_0$
such that
\begin{equation}
K_n \sim  c ~ \frac{\log n}{\gamma_1}
\label{eq:cond_for_initial_connec}
\end{equation}
for some $c>1$. Then we have
\begin{equation}
\lim_{n \to \infty} \bP{ C_{n, \gamma_1}(K_n) \cap \ldots \cap
C_{n,\gamma_\ell}(K_n) } = 1. \label{eq:continious_connectivity}
\end{equation}
} \label{thm:cntinuous_connectivity}
\end{theorem}

The event $[C_{\gamma_1,n}(K_n) \cap \ldots \cap
C_{\gamma_\ell,n}(K_n)]$ corresponds to the network in {\em each}
of its $\ell$ phases being connected as more nodes get added -- In
other words, on that event the sensors do form a connected network
at each phase of deployment. As a result, we infer via Theorem
\ref{thm:cntinuous_connectivity} that the condition
(\ref{eq:cond_for_initial_connec}) (with $c>1$) is enough to
ensure that the network remains a.a.s. connected as more sensors
are deployed over time.

The main conclusions of the paper, obtained by combining Theorem
\ref{thm:KeyRingSize} and Theorem
\ref{thm:cntinuous_connectivity}, can now be summarized as
follows:
\begin{corollary}
{\sl With $0 < \gamma_1 < \gamma_2 < \ldots < \gamma_{\ell} \leq 1
$, consider a scaling $K: \mathbb{N}_0 \rightarrow \mathbb{N}_0$
such that $K_n = O(\log n)$ with
\[
 K_n \geq \max
\left \{ ( \gamma_1)^{-1}, \lambda^\star \right \} \cdot \log n,
\quad n=2,3, \ldots \label{eq:ultimate}
\]
Then, the following holds:
\begin{enumerate}
\item The maximum number of keys kept in the memory module of each
sensor will be a.a.s. less than $3K_n$; \item The network deployed
gradually in $\ell$ steps (as in Section \ref{sec:ModelPairwise})
will be a.a.s. connected in each of the $\ell$ phases of
deployment.
\end{enumerate}
} \label{cor:ultimate}
\end{corollary}

\section{Simulation study}
\label{sec:Simulation}

\begin{figure}[!t]
\vspace{-.4cm} \centering\subfigure[]{\hspace{-0.5cm}
\includegraphics[totalheight=0.3\textheight,
width=0.48\textwidth]{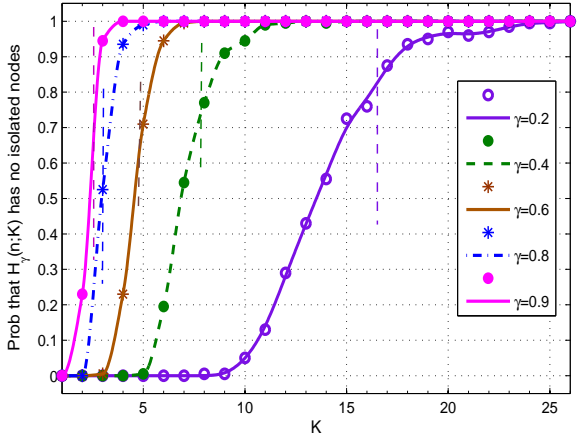} \label{figure1} } \subfigure[]
{\hspace{-0.5cm}
\includegraphics[totalheight=0.3\textheight,
width=0.48\textwidth]{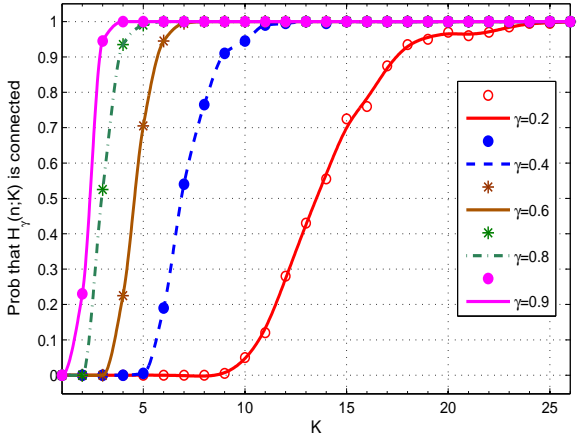} \label{figure2} }\caption{\sl{
$a)$ Probability that $\mathbb{H}_{\gamma}(n;K)$ contains no
isolated for $n=1000$; in each case, the empirical probability
value is obtained through $200$ experiments. Vertical dashed lines
stand for the critical thresholds asserted by Theorem
\ref{thm:Isol_Partial_Deployment}. It is clear that the
theoretical findings are in perfect agreement with the practical
observations. $b)$ Probability that $\mathbb{H}_{\gamma}(n;K)$ is
connected for $n=1,000$ obtained in the same way. Clearly, the
curves are almost indistinguishable from the corresponding ones of
part $(a)$; this supports the claim that absence of isolated nodes
and connectivity are asymptotically equivalent properties.}}
\vspace{-.5cm}
\end{figure}

We now present experimental results in support of the theoretical
findings. In each set of experiments, we fix $n$ and $\gamma$.
Then, we generate random graphs $\mathbb{H}_{\gamma}(n;K)$ for
each $K=1, \ldots , K_{\rm max}$ where the maximal value $K_{\rm
max}$ is selected large enough. In each case, we check whether the
generated random graph has isolated nodes and is connected. We
repeat the process $200$ times for each pair of values $\gamma$
and $K$ in order to estimate the probabilities of the events of
interest. For various values of $\gamma$, Figure \ref{figure1}
depicts the estimated probability $P^{\star}_\gamma (n; K)$ that
$\mathbb{H}_{\gamma}(n;K)$ has no isolated nodes as a function of
$K$. Here, $n$ is taken to be $1,000$. The plots in Figure
\ref{figure1} clearly confirm the claims of Theorem
\ref{thm:Isol_Partial_Deployment}: In each case $P^{\star}_\gamma
(n; K)$ exhibits a threshold behavior and the transitions from
$P^{\star}_\gamma (n; K)=0$ to $P^{\star}_\gamma (n; K)=1$ take
place around $K=r(\gamma)\frac{\log n}{\gamma}$ as dictated by
Theorem \ref{thm:Isol_Partial_Deployment}; the critical value
$K=r(\gamma)\frac{\log n}{\gamma}$ is shown by a vertical dashed
line in each plot.

Similarly, Figure \ref{figure2} shows the estimated probability
$P_\gamma (n; K)$ v.s. $K$ for various values of $\gamma$ with
$n=1000$. For each specified $\gamma$, we see that the variation
of $P_\gamma (n; K)$ with $K$ is almost indistinguishable from
that of $P^{\star}_\gamma (n; K)$ supporting the claim that
$P_\gamma (n; K)$ exhibits a full zero-one law similar to that of
Theorem \ref{thm:Isol_Partial_Deployment} with a threshold
behaving like $r(\gamma)$. We can also conclude by monotonicity
that $P_\gamma (n; K)=1$ whenever (\ref{eq:scaling_K}) holds with
$c>1$; this verifies Theorem \ref{thm:OneLaw_Partial_Deployment}.
Furthermore, it is evident from Figure \ref{figure2} that for a
given $K$ and $n$, $P_\gamma (n; K)$ increases as $\gamma$
increases supporting Theorem \ref{thm:cntinuous_connectivity}.


We also present experimental results that validate Lemma
\ref{lem:KeyRing} and Theorem \ref{thm:KeyRingSize}: For fixed
values of $n$ and $K$ we have constructed key rings according to
the mechanism presented in Section \ref{sec:ModelPairwise}. For
each pair of parameters $n$ and $K$, the experiments have been
repeated $1,000$ times yielding $1,000\times n$ key rings for each
parameter pair. The results are depicted in Figures 1-4 which show
the key ring sizes according to their frequency of occurrence. The
histograms in blue consider all of the produced $1,000 \times n$
key rings, while the histograms in white consider only the $1,000$
maximal key ring sizes, i.e., only the largest key ring among $n$
nodes in an experiment.

\begin{figure}[!t]
\centering\subfigure[]{\hspace{-0.5cm}
\includegraphics[totalheight=0.26\textheight,
width=0.5\textwidth]{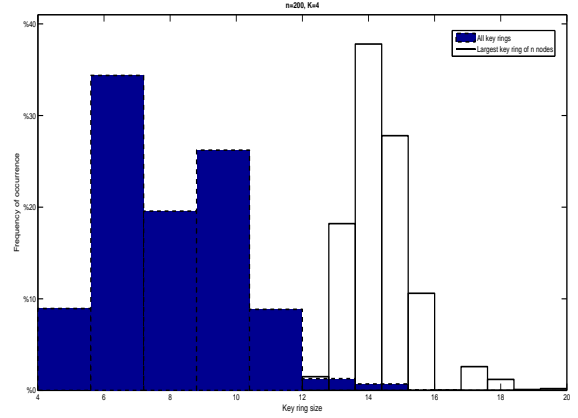} \label{fig:key_size_200} }
\subfigure[] {\hspace{-0.3cm}
\includegraphics[totalheight=0.26\textheight,
width=0.5\textwidth]{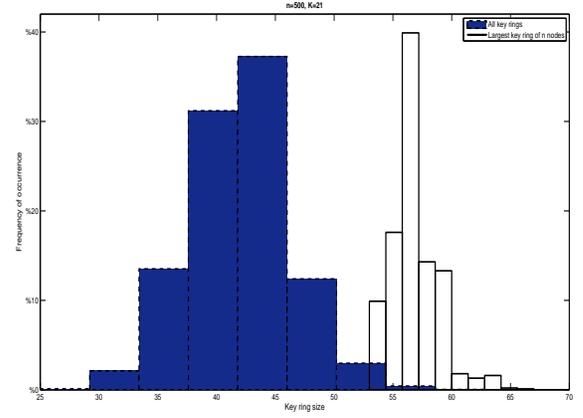} \label{fig:key_size_500} }
\caption[{\sl Key ring sizes for $n=200,K=4$ and
$n=500,K=21$.}]{\sl{ $a)$ Key ring sizes observed in $1,000$
experiments for $n=200$ and $K=4$ -- Only 2\% of the key rings are
larger than $3K$ and the largest key ring has size $20$. $b)$ Key
ring sizes observed in $1,000$ experiments for $n=500$ and $K=21$
-- Out of the $500,000$ key rings produced only $9$ happened to be
larger than $3K$ while the largest size observed is $67$.}}
\end{figure}

\begin{figure}[!h]
\centering\subfigure[]{\hspace{-0.5cm}
\includegraphics[totalheight=0.26\textheight,
width=0.5\textwidth]{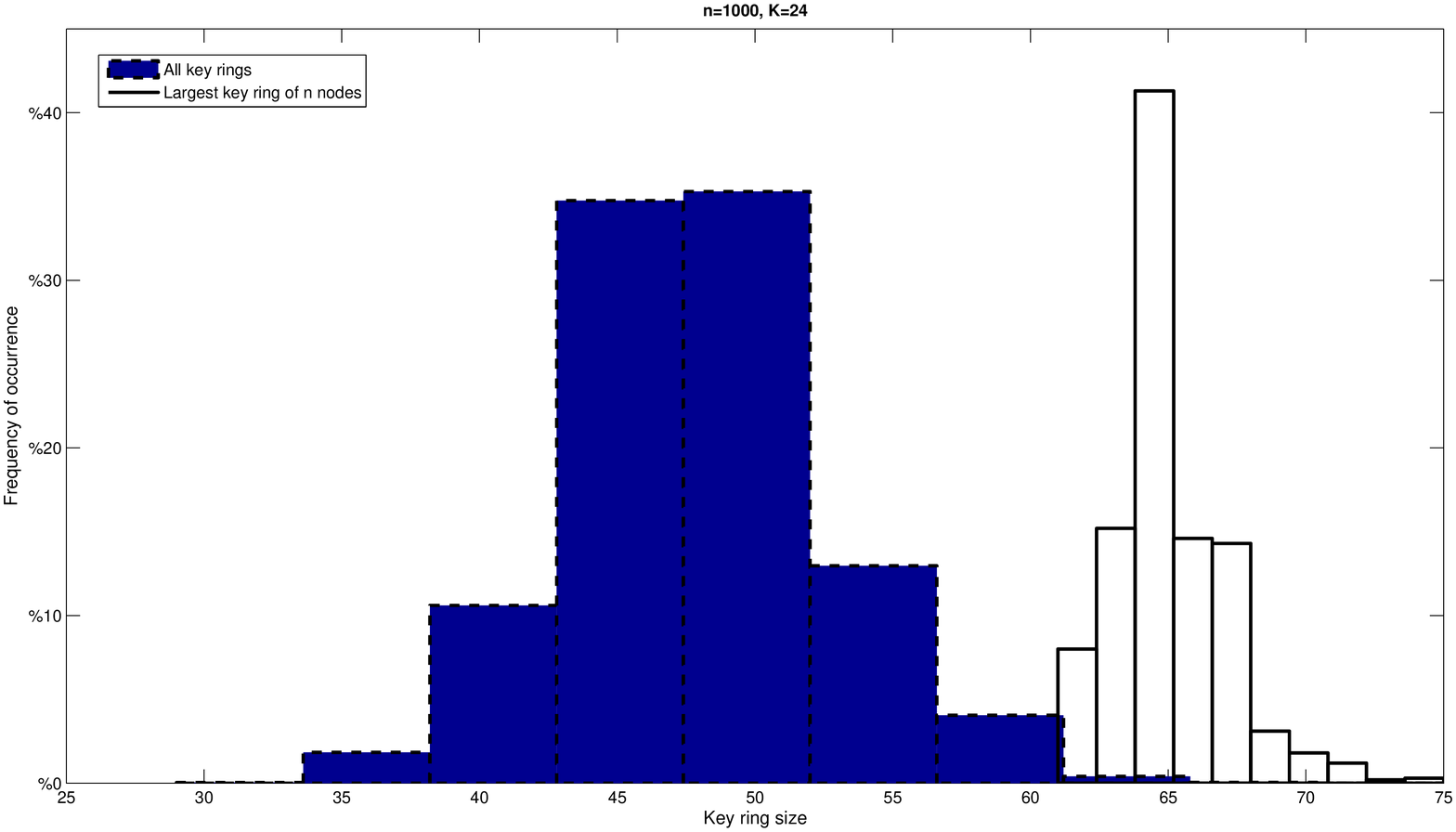} \label{fig:key_size_1000} }
\subfigure[] {\hspace{-0.3cm}
\includegraphics[totalheight=0.26\textheight,
width=0.5\textwidth]{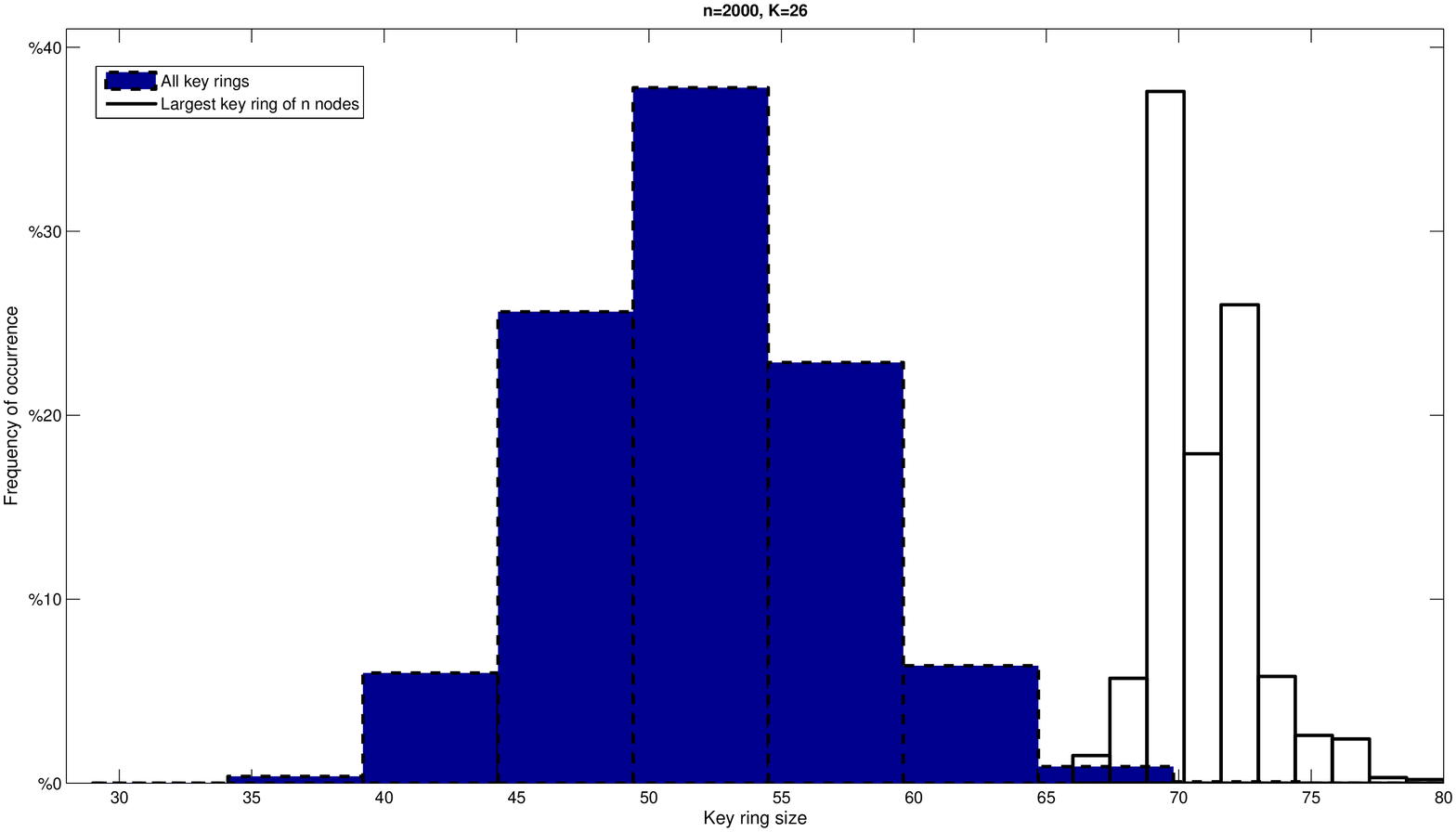} \label{fig:key_size_2000}
}\caption[{\sl Key ring sizes for $n=1000,K=24$ and
$n=2000,K=26$.}]{\sl{ $a)$ Key ring sizes observed in $1,000$
experiments for $n=1,000$ and $K=24$ -- $1,000,000$ key rings are
produced. Only $5$ of them happened to be larger than $3K$ and the
largest observed key ring size is $75$. $b)$ Key ring sizes
observed in $1,000$ experiments for $n=2,000$ and $K=26$ -- Out of
the 2000000 key rings produced only $2$ happened to be larger than
$3K$ the largest of them having $80$ keys.}}
\end{figure}

It is immediate from Figures
\ref{fig:key_size_200}-\ref{fig:key_size_2000} that the key ring
sizes tend to concentrate around $2K$, validating the claim of
Lemma \ref{lem:KeyRing}. As would be expected, this concentration
becomes more evident as $n$ gets large. It is also clear that, in
almost all cases the maximum size of a key ring (out of $n$ nodes)
is less than $3K$ validating the claim of Theorem
\ref{thm:KeyRingSize}.

\section{Conclusion}

In this paper, we consider the pairwise key distribution scheme of
Chan et al. which was proposed to establish security in wireless
sensor networks. This pairwise scheme has many advantages over
other key distribution schemes but deemed {\em not} scalable due
to $i)$ large number of keys required to establish secure
connectivity and $ii)$ the difficulties in the implementation when
sensors are required to be deployed in multiple stages. Here, we
address this issue and propose an implementation of the pairwise
scheme that supports the gradual deployment of sensor nodes in
several consecutive phases. We show how should the scheme
parameter be adjusted with the number $n$ of sensors so that the
secure connectivity can be maintained in the network throughout
all stages of the deployment. We also explore the relation between
the scheme parameter and the amount of memory that each sensor
needs to spare for storing secure keys. By showing that the
required number of keys is $O(\log n)$ to achieve connectivity at
every step of the deployment, we confirm the scalability of the
pairwise scheme in the context of WSNs.

\section{A proof of Theorem \ref{thm:OneLaw_Partial_Deployment}}
\label{sec:Proofs}

Fix $n=2,3, \ldots $ and $\gamma$ in the interval $(0,1)$, and
consider a positive integer $K \geq 2$. Throughout the discussion,
$n$ is sufficiently large so that the conditions
\begin{equation}
2(K+1) < n,
           \quad K+1 \leq n - \lfloor \gamma n \rfloor
           \quad \mbox{and} \quad 2 < \gamma n
\label{eq:OneLawConditions}
\end{equation}
are all enforced; these conditions are made in order to avoid
degenerate situations which have no bearing on the final result.
There is no loss of generality in doing so as we eventually let
$n$ go to infinity.

For any non-empty subset $R$ contained in $\{1, \ldots , \lfloor
\gamma n \rfloor \}$, we define the graph $\mathbb{H}_\gamma (n;K)
(R)$ (with vertex set $R$) as the subgraph of $\mathbb{H}_\gamma
(n;K)$ restricted to the nodes in $R$. We say that $R$ is {\em
isolated} in $\mathbb{H}_\gamma (n;K)$ if there are no edges (in
$\mathbb{H}_\gamma (n;K)$) between the nodes in $R$ and the nodes
in its complement $R^{c|\gamma} := \{ 1, \ldots , \lfloor \gamma n
\rfloor \} - R$. 
This is characterized by the event $B_{n,\gamma} (K ; R)$ given by
\[
B_{n,\gamma} (K; R) := \left[i \not \in \Gamma_{n,j}, j \notin
\Gamma_{n,i} ,
        \ i \in R , \ j \in R^{c|\gamma} \right] .
\]
Also, let $C_{n,\gamma} (K ;R)$ denote the event that the induced
subgraph $\mathbb{H}_\gamma (n;K) (R)$ is itself connected.
Finally, we set
\[
A_{n,\gamma}(K ; R) := C_{n,\gamma}(K ; R) \cap B_{n,\gamma} (K ;
R) .
\]

The discussion starts with the following basic observation: If
$\mathbb{H}_\gamma (n;K)$ is {\em not} connected, then there must
exist a non-empty subset $R$ of nodes contained in $\{1, \ldots ,
\lfloor \gamma n \rfloor \}$, such that $\mathbb{H}_\gamma(n;K)
(R)$ is itself connected while $R$ is isolated in
$\mathbb{H}_\gamma(n;K)$. This is captured by the inclusion
\begin{equation}
C_{n,\gamma} (K)^c \subseteq \cup_{R \in \mathcal{N}_{n,\gamma}} ~
A_{n,\gamma} (K ; R) \label{eq:BasicIdea}
\end{equation}
with $\mathcal{N}_{n,\gamma}$ denoting the collection of all
non-empty subsets of $\{ 1, \ldots , \lfloor \gamma n \rfloor \}$.
This union need only be taken over all non-empty subsets $R$ of
$\{1, \ldots , \lfloor  \gamma n \rfloor \}$ with $1 \leq |R| \leq
\lfloor \frac{ \lfloor \gamma n \rfloor }{2} \rfloor $, and it is
useful to note that $\lfloor \frac{ \lfloor \gamma n \rfloor }{2}
\rfloor = \lfloor \frac{ \gamma n }{2} \rfloor $. Then, a standard
union bound argument immediately gives
\begin{eqnarray}\nonumber
\bP{ C_{n,\gamma}(K)^c } &\leq& \sum_{ R \in
\mathcal{N}_{n,\gamma}: ~ 1 \leq |R| \leq \lfloor \frac{\gamma
n}{2} \rfloor } \bP{ A_{n, \gamma}(K ; R) }
\\
&=& \sum_{r=1}^{ \lfloor \frac{\gamma n}{2} \rfloor } \left (
\sum_{R \in \mathcal{N}_{n,\gamma,r} } \bP{ A_{n,\gamma}(K; R) }
\right ) \label{eq:BasicIdea+UnionBound}
\end{eqnarray}
where $\mathcal{N}_{n, \gamma, r}$ denotes the collection of all
subsets of $\{ 1, \ldots , \lfloor \gamma n \rfloor \}$ with
exactly $r$ elements.

For each $r=1, \ldots , \lfloor \gamma n \rfloor$, when $R = \{ 1,
\ldots , r \}$, we simplify the notation by writing $A_{n,\gamma,
r} (K) := A_{n,\gamma}(K ; R )$, $B_{n, \gamma , r} (K) :=
B_{n,\gamma}(K ; R )$ and $C_{n,\gamma, r} (K) := C_{n,\gamma} (K
; R )$. For $r=\lfloor \gamma n \rfloor $, the notation
$C_{n,\gamma, \lfloor \gamma n \rfloor}(K)$ coincides with
$C_{n,\gamma}(K)$ as defined earlier. Under the enforced
assumptions, it is a simple matter to check by exchangeability
that
\[
\bP{ A_{n,\gamma} (K ; R) } = \bP{ A_{n,\gamma , r} (K ) }, \quad
R \in \mathcal{N}_{n. \gamma, r}
\]
and the expression
\[
\sum_{R \in \mathcal{N}_{\gamma ,r}} \bP{ A_{n,\gamma} (K ; R) } =
{\lfloor \gamma n \rfloor \choose r} ~ \bP{ A_{n,\gamma, r}(K) }
\label{eq:ForEach=r}
\]
follows since $|\mathcal{N}_{n,\gamma, r} | = {\lfloor \gamma n
\rfloor \choose r}$. Substituting into
(\ref{eq:BasicIdea+UnionBound}) we obtain the bounds
\begin{eqnarray}
\bP{ C_{n,\gamma} (K)^c } \leq \sum_{r=1}^{ \lfloor \frac{\gamma
n}{2} \rfloor } {\lfloor \gamma n \rfloor \choose r} ~ \bP{
B_{n,\gamma,r}(K ) } \label{eq:BasicIdea+UnionBound2}
\end{eqnarray}
as we make use of the obvious inclusion $A_{n,\gamma, r}(K)
\subseteq B_{n,\gamma,r}(K)$. Under the enforced assumptions, we
get
\begin{eqnarray}
\lefteqn{ \bP{B_{n, \gamma,r}(K)} } & &
\label{eq:ProbCalculation} \\
&=& \left({ {n- \lfloor \gamma n \rfloor + r-1}\choose K }
        \over
      { {n-1} \choose K } \right)^{r}
\cdot \left( { {n-r-1}\choose K }
       \over { {n-1} \choose K }
\right)^{\lfloor \gamma n \rfloor -r}. \nonumber
\end{eqnarray}
To see why this last relation holds, recall that for the set $\{1,
\ldots , r \}$ to be isolated {\em in} $\mathbb{H}_\gamma (n;K)$
we need that (i) each of the nodes $r+1, \ldots , \lfloor \gamma n
\rfloor $ are adjacent only to nodes {\em outside} the set of
nodes $\{1, \ldots , r \}$; and (ii) none of the nodes $1, \ldots
, r$ are adjacent with any of the nodes $r+1, \ldots , \lfloor
\gamma n \rfloor $ -- This last requirement does not preclude
adjacency with any of the nodes $\lfloor \gamma n \rfloor + 1 ,
\ldots , n$. Reporting (\ref{eq:ProbCalculation}) into
(\ref{eq:BasicIdea+UnionBound2}), we conclude that
\begin{eqnarray}
\lefteqn{ \bP{ C_{n,\gamma}(K)^c } } & &
\label{eq:BasicIdea+UnionBound3} \\
&\leq&
\sum_{r=1}^{ \lfloor \frac{\gamma n}{2} \rfloor }
           {\lfloor \gamma n \rfloor \choose r}
\left( { {n- \lfloor \gamma n \rfloor +r-1}\choose K }
       \over { {n-1} \choose K } \right)^{r}
\cdot \left({ {n-r-1}\choose K }\over { {n-1} \choose K } \right)
^{\lfloor \gamma n \rfloor - r} \nonumber
\end{eqnarray}
with conditions (\ref{eq:OneLawConditions}) ensuring that the
binomial coefficients are well defined.

The remainder of the proof consists in bounding each of the terms
in (\ref{eq:BasicIdea+UnionBound3}). To do so we make use of
several standard bounds. First we recall the well-known bound
\[
{\lfloor \gamma n \rfloor \choose r} \leq \left(\frac{\lfloor
\gamma n \rfloor  e }{r}\right)^{r}, \quad r=1, \ldots , \lfloor
\gamma n \rfloor .
\]
Next, for $0\leq K \leq x \leq y$, we note that
\[
\frac{ {x \choose K} }{ {y \choose K} } = \prod_{\ell=0}^{K-1}
\left ( \frac{x-\ell}{y-\ell} \right ) \leq \left ( \frac{x}{y}
\right )^K \label{eq:UsefulBounds}
\]
since $\frac{x-\ell}{y-\ell}$ decreases as $\ell$ increases from
$\ell = 0$ to $\ell=K-1$.

Now pick $r=1, \ldots , \lfloor \gamma n \rfloor $. Under
(\ref{eq:OneLawConditions}) we can apply these bounds to obtain
\begin{eqnarray}
\lefteqn{ {\lfloor \gamma n \rfloor \choose r} \left( { {n-
\lfloor \gamma n \rfloor +r-1}\choose K }
         \over { {n-1} \choose K } \right)^{r}
\cdot \left( { {n-r-1}\choose K }\over { {n-1} \choose K }
\right)^{\lfloor \gamma n \rfloor - r} } & &
\nonumber \\
&\leq& \left(\frac{\lfloor \gamma n \rfloor  e }{r}\right)^{r}
\cdot \left ( \frac{n-\lfloor \gamma n \rfloor + r-1} {n-1} \right
)^{rK}
\nonumber \\
& & ~ \times \left ( \frac{n - r - 1} {n-1} \right )^{K(\lfloor
\gamma n \rfloor - r)}
\nonumber \\
&\leq& \left( \frac{\gamma n e}{r} \right)^{r} \left ( 1 -
\frac{\lfloor \gamma n \rfloor - r } {n-1} \right )^{rK} \left ( 1
- \frac{r} {n-1} \right )^{K (\lfloor \gamma n \rfloor - r)}
\nonumber \\
&\leq& \left(  \gamma n e \right)^{r} \cdot \left ( 1 -
\frac{\lfloor \gamma n \rfloor - r } {n} \right )^{rK} \cdot \left
( 1 - \frac{r} {n} \right )^{K (\lfloor \gamma n \rfloor - r)}
\nonumber \\
&\leq& \left( \gamma n e\right)^{r} \cdot e^{- \left (
\frac{\lfloor \gamma n \rfloor - r }{n} \right ) rK } \cdot e^{ -
\left ( \frac{r} {n} \right ) (\lfloor \gamma n \rfloor - r)K }.
\nonumber
\end{eqnarray}
It is plain that
\begin{eqnarray}
\bP{ C_{n,\gamma}(K)^c } &\leq& \sum_{r=1}^{ \lfloor \frac{\gamma
n}{2} \rfloor } \left(\gamma n e\right)^{r} \cdot e^{- 2 \left (
\frac{\lfloor \gamma n \rfloor - r }{n} \right ) rK }
\nonumber \\
&\leq& \sum_{r=1}^{ \lfloor \frac{\gamma n}{2} \rfloor } \left (
\gamma n e \cdot e^{- 2 \left ( \frac{\lfloor \gamma n \rfloor -
\lfloor \frac{\gamma n}{2} \rfloor }{n} \right ) K } \right )^r
\label{eq:BasicIdea+UnionBound4}
\end{eqnarray}
as we note that
\[
\frac{\lfloor \gamma n \rfloor - r }{n} \geq \frac{\lfloor \gamma
n \rfloor - \lfloor \frac{\gamma n}{2} \rfloor }{n} , \quad r=1,
\ldots , \lfloor \frac{\gamma n}{2} \rfloor .
\]

Next, consider a scaling $K: \mathbb{N}_0 \rightarrow
\mathbb{N}_0$ such that (\ref{eq:scaling_K}) holds for some $c>1$,
and replace $K$ by $K_n$ in (\ref{eq:BasicIdea+UnionBound4})
according to this scaling. Using the form (\ref{eq:scaling_K}) of
the scaling we get,
\[
a_n := \gamma n e \cdot e^{- 2 \left ( \frac{\lfloor \gamma n
\rfloor - \lfloor \frac{\gamma n}{2} \rfloor }{n} \right ) K_n } =
(\gamma e) \cdot n^{ 1 - 2 c_n \left ( \frac{\lfloor \gamma n
\rfloor - \lfloor \frac{\gamma n}{2} \rfloor }{\gamma n} \right )
}
\]
for each $n=1,2, \ldots$, with $\lim_{n \rightarrow \infty } c_n =
c$. It is a simple matter to check that
\[
\lim_{n \rightarrow \infty } \left ( 2 c_n \left ( \frac{\lfloor
\gamma n \rfloor - \lfloor \frac{\gamma n}{2} \rfloor }{\gamma n}
\right ) \right ) = c, \label{eq:LimitExponent}
\]
so that by virtue of the fact that $c > 1$, we have
\begin{equation}
\lim_{n \rightarrow \infty } a_n = 0. \label{eq:LimitExponent2}
\end{equation}

From (\ref{eq:BasicIdea+UnionBound4}) we conclude that
\[
\bP{ C_{n,\gamma}(K_n)^c } \leq \sum_{r=1}^{ \lfloor \frac{\gamma
n}{2} \rfloor } \left ( a_n \right )^r \leq \sum_{r=1}^{ \infty }
\left ( a_n \right )^r = \frac{a_n}{1- a_n}
\]
where for $n$ sufficiently large the summability of the geometric
series is guaranteed by (\ref{eq:LimitExponent2}). The conclusion
$\lim_{n \to \infty} \bP{ C_{n,\gamma}(K)^c } = 0$ is now a
straightforward consequence of the last bound, again by virtue of
(\ref{eq:LimitExponent2}).

\section{A proof of Theorem \ref{thm:Isol_Partial_Deployment}}
\label{sec:Proofs2}

Fix $n=2,3, \ldots $ and consider $\gamma$ in $(0,1)$ and positive
integer $K$ such that $K < n$. We write
\[
\chi_{n,\gamma,i}(K) := \1{ {\rm Node~}i~{\rm is~isolated~in~}
\mathbb{H}_{\gamma}(n;K) }
\]
for each $i=1, \ldots , \lfloor \gamma n \rfloor$. The number of
isolated nodes in $\mathbb{H}_{\gamma}(n; K)$ is simply given by
\[ I_{n,\gamma} (K) := \sum_{i=1}^{\lfloor \gamma
n \rfloor} \chi_{n,\gamma, i}(K),
\]
whence the random graph $\mathbb{H}_{\gamma}(n; K)$ has no
isolated nodes if $I_{n,\gamma} (K) = 0$. The method of first
moment \cite[Eqn (3.10), p. 55]{JansonLuczakRucinski} and second
moment \cite[Remark 3.1, p. 55]{JansonLuczakRucinski} yield the
useful bounds
\begin{equation}
1 - \bE{ I_{n,\gamma} (K) } \leq \bP{  I_{n,\gamma} (K) = 0 } \leq
1 - \frac{ \bE{ I_{n,\gamma} (K)}^2}{ \bE{ I_{n,\gamma} (K) ^2} }.
\label{eq:FirstMoment}
\end{equation}

The rvs $\chi_{n,\gamma,1}(K), \ldots , \chi_{n,\gamma,\lfloor
\gamma n \rfloor} (K)$ being exchangeable, we find
\begin{equation}
\bE{ I_{n,\gamma} (K)} = \lfloor \gamma n \rfloor \bE{
\chi_{n,\gamma,1}(K) } \label{eq:FirstMomentExpression}
\end{equation}
and
\begin{eqnarray}
\lefteqn{ \bE{ I_{n,\gamma} (K)^2 } } & &
\nonumber \\
&=& \lfloor \gamma n \rfloor \bE{ \chi_{n,\gamma,1}(K) }
\label{eq:SecondMomentExpression} \\
& & ~ + \lfloor \gamma n \rfloor (\lfloor \gamma n \rfloor -1)
\bE{ \chi_{n,\gamma,1}(K) \chi_{n,\gamma,2}(K) } \nonumber
\end{eqnarray}
by the binary nature of the rvs involved. It then follows in the
usual manner that
\begin{eqnarray}
\frac{ \bE{ I_{n,\gamma} (K)^2 }}{ \bE{ I_{n,\gamma} (K) }^2 } &=&
\frac{1}{\lfloor \gamma n \rfloor \bE{ \chi_{n,\gamma,1}(K) } }
\label{eq:SecondMomentRatio} \\
& & ~ + \frac{ \lfloor \gamma n \rfloor -1}{ \lfloor \gamma n
\rfloor } \frac{\bE{ \chi_{n,\gamma,1}(K) \chi_{n,\gamma,2}(K) }}
     {\left (  \bE{ \chi_{n,\gamma,1}(K) } \right )^2 } .
\nonumber
\end{eqnarray}

From (\ref{eq:FirstMoment}) and (\ref{eq:FirstMomentExpression})
we conclude that the one-law $\lim_{n\to \infty}
\bP{I_{n,\gamma}(K_n) = 0} = 1$ holds if we show that
\begin{equation}
\lim_{n \to \infty} \lfloor \gamma n \rfloor  \bE{
\chi_{n,\gamma,1} (K_n) }= 0.
\label{eq:OneLaw+NodeIsolation+convergence}
\end{equation}
On the other hand, it is plain from (\ref{eq:FirstMoment}) and
(\ref{eq:SecondMomentRatio}) that the zero-law $\lim_{n \to
\infty} \bP{I_{n,\gamma}(K_n) = 0} = 0$ will be established if
\begin{equation}
\lim_{n \to \infty} \lfloor \gamma n \rfloor \bE{ \chi_{n,1} (K_n)
}= \infty \label{eq:OneLaw+NodeIsolation+convergence2}
\end{equation}
and
\begin{equation}
\limsup_{n \to \infty} \left(\frac{\bE{ \chi_{n,\gamma,1} (K_n)
\chi_{n,\gamma,2} (K_n) }}
     {\left (  \bE{ \chi_{n,\gamma,1} (K_n) } \right )^2 }\right)
\leq 1. \label{eq:ZeroLaw+NodeIsolation+convergence}
\end{equation}

The next two technical lemmas establish
(\ref{eq:OneLaw+NodeIsolation+convergence}),
(\ref{eq:OneLaw+NodeIsolation+convergence2}) and
(\ref{eq:ZeroLaw+NodeIsolation+convergence}) under the appropriate
conditions on the scaling $K: \mathbb{N}_0 \rightarrow
\mathbb{N}_0$.

\begin{lemma}
{\sl Consider $\gamma$ in $(0,1)$ and a scaling $K: \mathbb{N}_0
\rightarrow \mathbb{N}_0$ such that (\ref{eq:scaling_K}) holds for
some $c>0$. We have
\begin{equation}
\lim_{n \rightarrow \infty } n\bE{ \chi_{n,\gamma, 1} (K_n) } =
\left \{
\begin{array}{ll}
0 & \mbox{if~ $c > r(\gamma)$} \\
  &                      \\
\infty & \mbox{if~$c < r(\gamma)$}
\end{array}
\right . \label{eq:NodeIsolation+FirstMoment}
\end{equation}
with $r(\gamma)$ specified via (\ref{eq:r_gamma_defn}).
 }
\label{lem:Technical1}
\end{lemma}

\begin{lemma}
{\sl Consider $\gamma$ in $(0,1)$ and a scaling $K: \mathbb{N}_0
\rightarrow \mathbb{N}_0$ such that (\ref{eq:scaling_K}) holds for
some $c>0$. We have
\begin{equation}
\limsup_{n \to \infty} \left( \frac{\bE{ \chi_{n,\gamma,1} (K_n)
\chi_{n,\gamma,2} (K_n) }}
     {\left (\bE{ \chi_{n,\gamma,1} (K_n) } \right )^2 }
\right ) \leq 1.
      \label{eq:Technical2}
\end{equation}
\label{lem:Technical2} }
\end{lemma}
Proofs of Lemma \ref{lem:Technical1} and Lemma
\ref{lem:Technical2} can be found in Section
\ref{subsec:ProofLemmaTechnical1} and Section
\ref{subsec:ProofLemmaTechnical2}, respectively. To complete the
proof of Theorem \ref{thm:Isol_Partial_Deployment}, pick a scaling
$K: \mathbb{N}_0 \rightarrow \mathbb{N}_0$ such that
(\ref{eq:scaling_K}) holds for some $c>0$. Under the condition $c>
r(\gamma)$ we get (\ref{eq:OneLaw+NodeIsolation+convergence}) from
Lemma \ref{lem:Technical1} and the one-law $\lim_{n\to \infty}
\bP{I_{n,\gamma}(K_n) = 0} = 1$ follows. Next, assume the
condition $c< r(\gamma)$. We obtain
(\ref{eq:OneLaw+NodeIsolation+convergence2}) and
(\ref{eq:ZeroLaw+NodeIsolation+convergence}) with the help of
Lemmas \ref{lem:Technical1} and \ref{lem:Technical2},
respectively, and the conclusion $\lim_{n \to \infty}
\bP{I_{n,\gamma}(K_n) = 0} = 0$ is now immediate.


\subsection{A proof of Lemma \ref{lem:Technical1}}
\label{subsec:ProofLemmaTechnical1}

Fix $n=2,3, \ldots $ and $\gamma$ in $(0,1)$, and consider a
positive integer $K$ such that $K < n$. Here as well there is no
loss of generality in assuming $n- \lfloor \gamma n \rfloor \geq K
$ and $\lfloor \gamma n \rfloor > 1$. Under the enforced
assumptions, we get
\begin{eqnarray}
\bE{ \chi_{n,\gamma, 1} (K) }
 &=& \frac{ {n- \lfloor
\gamma n \rfloor \choose K} }
       { {n- 1 \choose K} }
\left( \frac{{n- 2 \choose K}} {{n- 1 \choose K}} \right)
^{\lfloor \gamma n \rfloor -1}
\nonumber \\
&=& a(n;K) \cdot \left( 1- \frac{K}{n-1} \right)^{ \lfloor \gamma
n \rfloor -1} \label{eq:FirstMomentComputation}
\end{eqnarray}
with
\[
a(n;K) := \frac{(n- \lfloor \gamma n \rfloor)! }
     {(n- \lfloor \gamma n \rfloor-K)!}
\cdot \frac{(n-1-K)!}
     {(n-1)!} .
\]

Now pick a scaling $K: \mathbb{N}_0 \rightarrow \mathbb{N}_0$ such
that (\ref{eq:scaling_K}) holds for some $c>0$ and replace $K$ by
$K_n$ in (\ref{eq:FirstMomentComputation}) with respect to this
scaling. Applying Stirling's formula
\[
m! \sim \sqrt{2 \pi m} \left(\frac{m}{e}\right)^{m} \quad (m
\rightarrow \infty)
\]
to the factorials appearing in (\ref{eq:FirstMomentComputation}),
we readily get
\begin{eqnarray}
a(n;K_n) &\sim& \sqrt{ \frac{(n- \lfloor \gamma n \rfloor
)(n-1-K_n)}
     {(n- \lfloor \gamma n \rfloor-K_n)(n-1)}}
\cdot \alpha_n \beta_n
\nonumber \\
&\sim& \alpha_n \beta_n \label{eq:lem_tech_1_eq}
\end{eqnarray}
under the enforced assumptions on the scaling with
\begin{eqnarray}
\alpha_n &:=& \frac{(n-K_n-1)^{n-K_n-1}}{(n-1)^{n-1}}
\nonumber \\
&=& \left( 1-\frac{K_n}{n-1} \right)^{n-1} \cdot \left ( n-K_n-1
\right )^{-K_n} \nonumber
\end{eqnarray}
and
\begin{eqnarray}
\beta_n &:=& \frac{(n- \lfloor \gamma n \rfloor )^{n- \lfloor
\gamma n \rfloor }}
     {(n- \lfloor \gamma n \rfloor-K_n)^{n- \lfloor \gamma n \rfloor-K_n}}
\nonumber \\
&=& \left( 1-\frac{K_n}{n-\lfloor \gamma n \rfloor }
\right)^{-(n-\lfloor \gamma n \rfloor )} \cdot \left ( n- \lfloor
\gamma n \rfloor-K_n \right )^{K_n} . \nonumber
\end{eqnarray}

In obtaining the asymptotic behavior of (\ref{eq:lem_tech_1_eq})
we rely on the following technical fact: For any sequence $m:
\mathbb{N}_0 \rightarrow \mathbb{N}_0$ with $m_n = O(n)$, we have
\begin{equation}
\left( 1 - \frac{K_n}{m_n} \right)^{m_n} \sim e^{ - K_n}.
\label{eq:m_n}
\end{equation}
To see why (\ref{eq:m_n}) holds, recall the elementary
decomposition
\[
\log (1-x) = - x - \Psi(x) \ \mbox{with} \quad \Psi (x) :=
\int_0^x \frac{t}{1-t} ~dt
\]
valid for $0 \leq x < 1$. Using this fact, we get
\begin{equation}
\left( 1 - \frac{K_n}{m_n} \right)^{m_n} = e^{ - K_n} \cdot e^{ -
m_n \Psi \left ( \frac{K_n}{m_n} \right ) } \label{eq:m_n+A}
\end{equation}
for all $n=1,2, \ldots$.

Under the enforced assumptions we have $m_n = O(n)$ and $K_n =
O(\log n)$, so that
\[
\lim_{n \rightarrow \infty} \frac{K_n}{m_n} = 0 \quad \mbox{and}
\quad \lim_{n \rightarrow \infty} m_n \left ( \frac{K_n}{m_n}
\right )^2 = 0.
\]
It is now plain that
\[
\lim_{n \rightarrow \infty} m_n \Psi \left ( \frac{K_n}{m_n}
\right ) = 0
\]
as we note that $\lim_{x \downarrow 0} \frac{\Psi(x)}{x^2} =
\frac{1}{2}$. This establishes (\ref{eq:m_n}) via
(\ref{eq:m_n+A}).

Using (\ref{eq:m_n}), first with $m_n = n-1$, then with $m_n =
n-\lfloor \gamma n \rfloor$, we obtain
\[
\left( 1-\frac{K_n}{n-1} \right)^{n-1} \sim e^{-K_n}
\]
and
\[
\left( 1-\frac{K_n}{n-\lfloor \gamma n \rfloor }
\right)^{-(n-\lfloor \gamma n \rfloor)} \sim \left ( e^{ - K_n }
\right )^{-1} = e^{ K_n },
\]
whence
\begin{equation}
\alpha_n \beta_n \sim \left ( \frac{ n- \lfloor \gamma n \rfloor -
K_n }
     { n - K_n-1 }
\right )^{K_n} .
\end{equation}
With the help of (\ref{eq:FirstMomentComputation}) and
(\ref{eq:lem_tech_1_eq}) we now conclude that
\begin{eqnarray}
\lefteqn{ n \bE{ \chi_{n,\gamma, 1} (K_n) } } & &
\label{eq:Intermediary1} \\
&\sim& n \left( 1- \frac{K_n}{n-1} \right)^{ \lfloor \gamma n
\rfloor -1} \cdot \left ( \frac{ n- \lfloor \gamma n \rfloor - K_n
}
     { n - K_n-1 }
\right )^{K_n}. \nonumber
\end{eqnarray}

A final application of (\ref{eq:m_n}), this time with $m_n = n-1$,
gives
\begin{eqnarray}
\left( 1- \frac{K_n}{n-1} \right)^{ \lfloor \gamma n \rfloor -1}
&=& \left ( \left( 1- \frac{K_n}{n-1} \right)^{n-1} \right )^{
\frac{\lfloor \gamma n \rfloor -1}{n-1} }
\nonumber \\
& \sim & e^{ - \frac{\lfloor \gamma n \rfloor -1}{n-1} K_n }
\label{eq:Intermediary2}
\end{eqnarray}
since $ \lim_{n \rightarrow \infty} \frac{\lfloor \gamma n \rfloor
-1}{n-1} = \gamma $. Reporting (\ref{eq:Intermediary2}) into
(\ref{eq:Intermediary1}) we obtain
\begin{equation}
n \bE{ \chi_{n,\gamma, 1} (K_n) } \sim  e^{ \zeta_n }
\label{eq:ExponentialZeta}
\end{equation}
with
\[
\zeta_n := \log n - \left ( \frac{\lfloor \gamma n \rfloor
-1}{n-1} + \log \left ( \frac{ n- \lfloor \gamma n \rfloor - K_n }
     { n - K_n-1 }
\right ) \right ) K_n
\]
for all $n=1,2, \ldots $. Finally, from the condition
(\ref{eq:scaling_K}) on the scaling, we see that
\[
\lim_{n \rightarrow \infty} \frac{ \zeta_n}{\log n} = 1 - c + c
\frac{ \log (1-\gamma) }{ \gamma } = 1 - \frac{c}{r(\gamma) }.
\]
Thus, $\lim_{n \rightarrow \infty} \zeta_n = -\infty$ (resp.
$\infty$) if $r(\gamma) > c$ (resp. $r(\gamma) < c$) and the
desired result follows upon using (\ref{eq:ExponentialZeta}).

\subsection{A proof of Lemma \ref{lem:Technical2}}
\label{subsec:ProofLemmaTechnical2}

Fix positive integers $n=3,4, \ldots $ and $K$ with $K<n$. With
$\gamma$ in $(0,1)$, we again assume that $n- \lfloor \gamma n
\rfloor \geq K $ and $\lfloor \gamma n \rfloor > 1$. It is a
simple matter to check that
\[
\bE{ \chi_{n,\gamma,1}(K) \chi_{n,\gamma,2}(K) } = \left( { {n-
\lfloor \gamma n \rfloor } \choose { K} } \over { {n- 1 } \choose
{ K} } \right)^{2} \left( { {n- 3 } \choose { K} } \over { {n- 1 }
\choose { K} } \right)^{ \lfloor \gamma n \rfloor -2}
\]
and invoking (\ref{eq:FirstMomentComputation}) we readily conclude
that
\begin{eqnarray}\nonumber
\lefteqn{\frac{\bE{ \chi_{n,\gamma,1} (K) \chi_{n,\gamma,2} (K) }}
     {\left (  \bE{ \chi_{n,\gamma,1} (K) } \right )^2 }} &&
     \\
&=& \left( { {n- 3 } \choose { K} } \over { {n- 1 } \choose { K} }
\right)^{ \lfloor \gamma n \rfloor -2} \cdot \left( \frac{{n- 1
\choose K}}
     {{n- 2 \choose K}}
\right)^{2\left ( \lfloor \gamma n \rfloor -1 \right )}
\nonumber \\
&=& \left ( \left ( \frac{n-1-K}{n-1} \right ) \left (
\frac{n-2-K}{n-2} \right ) \right)^{ \lfloor \gamma n \rfloor -2}
\nonumber \\
& & ~ \times
 \left ( \frac{n-1}{n-1-K} \right ) ^{2\left ( \lfloor
\gamma n \rfloor -1 \right )}
\nonumber \\
&=& \left ( \frac{n-2-K}{n-2} \right )^{\lfloor \gamma n \rfloor
-2} \cdot \left ( \frac{n-1}{n-1-K} \right )^{\lfloor \gamma n
\rfloor }
\nonumber \\
&=& \left ( 1 - \frac{K}{n-2} \right )^{\lfloor \gamma n \rfloor
-2} \cdot \left ( 1 + \frac{K}{n-1-K} \right )^{\lfloor \gamma n
\rfloor }
\nonumber \\
&\leq& e^{ - K \cdot E(n;K) } \label{eq:KeyInequality}
\end{eqnarray}
where we have set
\[
E(n;K) := \frac{ \lfloor \gamma n \rfloor -2 }
       {n-2}
- \frac{ \lfloor \gamma n \rfloor }
     { n-1-K} .
\]
Elementary calculations show that
\[
- K \cdot E(n;K) = \frac{\lfloor \gamma n \rfloor}{n-2} \cdot
\frac{ K(K-1)}{ n-1-K} + \frac{2K}{n-2}.
\]

Now pick a scaling $K: \mathbb{N}_0 \rightarrow \mathbb{N}_0$ such
that (\ref{eq:scaling_K}) holds for some $c>0$. It is plain that
$\lim_{n \rightarrow \infty} K_n E(n;K_n) = 0$ and the conclusion
(\ref{eq:Technical2}) follows from (\ref{eq:KeyInequality}).

\section{A proof of Theorem \ref{thm:cntinuous_connectivity}}

Pick $0 < \gamma_1 < \gamma_2 < \ldots < \gamma_{\ell} \leq 1 $
and consider a scaling $K: \mathbb{N}_0 \rightarrow \mathbb{N}_0$
such that
\[
K_n \sim  c ~ \frac{\log n}{\gamma_1}
\]
for some $c>1$. It is plain that
(\ref{eq:continious_connectivity}) will hold provided
\begin{equation}
\lim_{n \to \infty} \bP{C_{n,\gamma_k }(K_n)} = 1, \quad k=1,
\ldots , \ell . \label{eq:continious_connectivityB}
\end{equation}
For each $k=1,2, \ldots , \ell$, we note that
\[
c ~ \frac{\log n}{\gamma_1} = c_k ~ \frac{\log n}{\gamma_k} \quad
\mbox{with~} c_k := c ~ \frac{\gamma_k}{\gamma_1}
\]
for all $n=1,2, \ldots $. But $c>1$ implies $c_k > 1$ since
$\gamma_1 < \ldots < \gamma_\ell$. As a result,
$\mathbb{H}_{\gamma_k}(n;K_n)$ will be a.a.s. connected by virtue
of Theorem \ref{thm:OneLaw_Partial_Deployment} applied to
$\mathbb{H}_{\gamma_k}(n; K)$, and
(\ref{eq:continious_connectivityB}) indeed holds.

\section{A proof of Lemma \ref{lem:KeyRing}}
\label{sec:KingRingSizesMainResults}

Fix $n=2,3, \ldots$ and positive integer $K$ with $K < n$. For
each $i=1,2, \ldots , n$, node $i$ is assigned a key ring
$\Sigma_{n,i}$ whose size is given by
\begin{equation}
|\Sigma_{n,i}| = | \Gamma_{n,i}| + \sum_{j=1, \ j \neq i}^n \1{ i
\in \Gamma_{n,j} } . \label{eq:SizeOfSigma}
\end{equation}
This is a simple consequence of the definition
(\ref{eq:KeyRingDefn}). We also define the maximal key ring size
as
\[
M_n := {\max}_{i=1, \ldots , n} |\Sigma_{n,i}| .
\]

It is easy to see that
\begin{equation}
|\Sigma_{n,i}| = K + B_{n,i} \label{eq:FromStoB}
\end{equation}
where $B_{n,i}$ is the rv determined through
\[
B_{n,i} := \sum_{j=1, \ j \neq i}^n \1{ i \in \Gamma_{n,j} }.
\]
Under the enforced independence assumptions, the rv $B_{n,i}$ is a
binomial rv ${\rm Bin}(n-1, \frac{K}{n-1})$, with
\[
\bE{ B_{n,i} } = (n-1) \cdot \frac{K}{n-1} = K
\]
and
\[
{\rm Var} [ B_{n,i} ] = (n-1) \cdot \frac{K}{n-1} \cdot \frac{n-1
- K }{n-1}.
\]

As a result, $\bE{ | \Sigma_{n,i} | } = 2 K$ and
\[
{\rm Var} [ |\Sigma_{n,i}| ] = K \left ( 1 - \frac{ K }{n-1}
\right ).
\]
It is now plain that
\begin{eqnarray}
\bE{ \left | \frac{|\Sigma_{n,i}|}{\bE{|\Sigma_{n,i}|}} - 1 \right
|^2 } &=& \frac{ {\rm Var} [ | \Sigma_{n,i} | ] }{
\bE{|\Sigma_{n,i}|}^2 }
\nonumber \\
&=& \frac{1}{4} \left ( \frac{1}{K} - \frac{ 1 }{n-1} \right )
\end{eqnarray}
so that
\begin{equation}
\bE{ \left | \frac{|\Sigma_{n,i}|}{2K} - 1 \right |^2 } =
\frac{1}{4} \left ( \frac{1}{K} - \frac{ 1 }{n-1} \right ) .
\label{eq:RatioVarToMeanSquared}
\end{equation}

 Under the enforced assumptions, we have
\[
\lim_{n \rightarrow \infty} \bE{ \left |
\frac{|\Sigma_{n,1}(K_n)|}{2K_n} - 1 \right |^2 } = 0
\]
by the earlier calculations (\ref{eq:RatioVarToMeanSquared}), and
the desired result (\ref{eq:keyring_lemma}) follows. \myendpf

\section{A proof of Theorem \ref{thm:KeyRingSize}}
\label{sec:ProofTheoremKeyRingSize}

Fix the positive integers $n=2,3, \ldots $ and $K$ with $K < n$.
Using (\ref{eq:FromStoB}) we readily get
\[
\left ( \max_{i=1, \ldots , n } |\Sigma_{n,i}| \right ) - 2 K =
\max_{i=1, \ldots , n } \left ( B_{n,i} - K \right ) .
\]
Therefore, with any given $t > 0$, we find
\begin{eqnarray}
\lefteqn{ \bP{ \left | \left ( \max_{i=1, \ldots , n}
|\Sigma_{n,i}| \right ) - 2 K \right | > t } } & &
\nonumber \\
&=& \bP{ \left | \max_{i=1, \ldots , n } \left ( B_{n,i} - K
\right ) \right |
> t }
\nonumber \\
&=& \bP{ \max_{i=1, \ldots , n } B_{n,i}
> K + t }
\nonumber \\
& & ~+ \bP{ \max_{i=1, \ldots , n } B_{n,i} < K - t } .
\label{eq:BoundForConcentrationResult}
\end{eqnarray}

We take each term in turn. First a simple union argument shows
that
\begin{eqnarray}
\lefteqn{ \bP{ {\max}_{i=1, \ldots , n} B_{n,i} > K+t } } & &
\nonumber \\
&=& \bP{ \cup_{i=1}^n [ B_{n,i} > K+t ] }
\nonumber \\
&\leq & \sum_{i=1}^n \bP{ B_{n,i} > K+t }
\nonumber \\
&=& n \bP{ B_{n,1} > K+t } \label{eq:KeyRingInequality1}
\end{eqnarray}
since the rvs $B_{n,1}, \ldots , B_{n,n}$ are identically
distributed (but not independent). Next we note that
\begin{eqnarray}
\lefteqn{ \bP{ {\max}_{i=1, \ldots , n} B_{n,i} < K-t } } & &
\nonumber \\
&=& \bP{ B_{n,i} < K-t, \ i=1, \ldots n }
\nonumber \\
&\leq& {\min}_{i=1, \ldots , n} \bP{ B_{n,i} <  K-t }
\nonumber \\
&=& \bP{ B_{n,1} <  K_n-t }. \label{eq:KeyRingInequality2}
\end{eqnarray}

To proceed we recall standard bounds for the tails of binomial rvs
\cite[lemma 1.1, p. 16]{PenroseBook}: With
\[
H(t) := 1 -t  + t \log t,
\]
we have the concentration inequalities
\[
\bP{ B_{n.1} > K +t } \leq e^{ - K \cdot H( \frac{ K + t }{ K } )
}
\]
and
\[
\bP{ B_{n,1} < K-t } \leq e^{ - K \cdot H( \frac{ K - t }{ K } ) }
\]
where the additional condition $0 < t < K$ is required for the
second inequality to hold. Simple calculations on the appropriate
ranges show that
\[
- K \cdot H \left ( \frac{ K \pm t }{ K } \right ) = \pm t - \left
( K \pm t \right ) \cdot \log \left ( 1 \pm \frac{ t }{ K } \right
) .
\]

Thus, by the first concentration inequality, we conclude from
(\ref{eq:KeyRingInequality1}) that
\begin{equation}
\bP{ {\max}_{i=1, \ldots , n} B_{n,i} > K+t } \leq e^{A_n(K;t)}
\label{eq:ConsequenceConcentration1}
\end{equation}
with
\[
A_n (K;t) := \log n + t - \left ( K + t \right ) \cdot \log \left
( 1 + \frac{ t }{ K }  \right ) .
\]
The second concentration inequality and
(\ref{eq:KeyRingInequality2}) together yield
\begin{equation}
\bP{ {\max}_{i=1, \ldots , n} B_{n,i} < K-t } \leq e^{B_n(K;t)}
\label{eq:ConsequenceConcentration2}
\end{equation}
with
\[
B_n (K;t) := -t -  \left ( K - t \right ) \cdot \log \left ( 1 -
\frac{ t }{ K }  \right )
\]
under the additional constraint $0 < t < K$.

Now consider a scaling $K: \mathbb{N}_0 \rightarrow \mathbb{N}_0$
of the form (\ref{eq:FormOfScaling}) for some $\gamma > 0$, and
select the sequence $t: \mathbb{N}_0 \rightarrow \mathbb{R}_+$
given by
\[
t_n = c log n, \quad n=1,2, \ldots
\]
with $c$ in the interval $(0,\gamma)$ (so that $0 < t_n < K_n$ for
all $n$ sufficiently large).

 Under appropriate
conditions on $\gamma$ and $c$, we shall show that
\begin{equation}
\lim_{n \rightarrow \infty } A_n (K_n; t_n) = -\infty
\label{eq:LimA}
\end{equation}
and
\begin{equation}
\lim_{n \rightarrow \infty } B_n (K_n; t_n) = -\infty.
\label{eq:LimB}
\end{equation}
The convergence statements
\[
\lim_{n \rightarrow \infty } \bP{ {\max}_{i=1, \ldots , n}
B_{n,i}(K_n) > K_n+t_n } = 0
\]
and
\[
\lim_{n \rightarrow \infty } \bP{ {\max}_{i=1, \ldots , n}
B_{n,i}(K_n) < K_n-t_n } = 0
\]
then follow from (\ref{eq:ConsequenceConcentration1}) and
(\ref{eq:ConsequenceConcentration2}), respectively, and the
desired conclusion (\ref{eq:KeyRingSize}) flows from
(\ref{eq:BoundForConcentrationResult}).

With the selections made above, we get $A_n (K_n; t_n) \sim
a(\gamma;c) \log n$ and $B_n (K_n; t_n) \sim b(\gamma;c) \log n$
with coefficients $a(\gamma;c)$ and $b(\gamma;c)$ given by
\[
a(\gamma;c) := 1 + c - \left ( \gamma + c \right ) \cdot \log
\left ( 1 + \frac{c}{\gamma} \right ) , \quad c > 0
\]
and
\[
b(\gamma;c) := - c - \left ( \gamma - c \right ) \cdot \log \left
( 1 - \frac{c}{\gamma} \right ) , \quad 0 < c < \gamma .
\]
Thus, in order to ensure (\ref{eq:LimA}) and (\ref{eq:LimB}), we
need to find $c$ in the interval $(0,\gamma)$ such that
$a(\gamma;c) < 0 $ and $b(\gamma;c) < 0 $, respectively. To that
end, we first note that
\[
\frac{\partial a}{\partial c} (\gamma ; c ) = - \log \left ( 1 +
\frac{c}{\gamma} \right ) < 0, \quad c > 0
\]
and
\[
\frac{\partial b}{\partial c} (\gamma ; c ) = \log \left ( 1 -
\frac{c}{\gamma} \right ) < 0, \quad 0 < c < \gamma .
\]
Therefore, both mappings $c \rightarrow a(\gamma;c)$ and $c
\rightarrow b(\gamma;c)$ are {\em strictly} decreasing on the
intervals $(0, \infty)$ and $(0,\gamma )$, respectively. Since
$\lim_{c \downarrow 0 } b(\gamma;c) = 0$, it is plain that
$b(\gamma;c) < 0 $ on the entire interval $(0,\gamma )$. On the
other hand, it is easy to check that $ \lim_{c \downarrow 0 }
a(\gamma;c) = 1$ and
\[
\lim_{c \uparrow \gamma } a(\gamma;c) = 1 -  \gamma \left ( 2 \log
2 - 1 \right ) = 1 - \frac{\gamma}{\gamma^\star}.
\]
Hence, if we select $\gamma > \gamma^\star$, then $a(\gamma;c) <
0$ for all $ c > c(\gamma)$ where $c(\gamma)$ is the unique
solution to the equation
\begin{equation}
a(\gamma ; c ) = 0, \quad c > 0. \label{eq:Equation1}
\end{equation}
Uniqueness is a consequence of the strict monotonicity mentioned
earlier.

The proof will be completed by showing that the constraint
\begin{equation}
c(\gamma) < \gamma, \quad \gamma > \gamma^\star
\label{eq:NeededInequality}
\end{equation}
indeed holds. For each $\gamma > 0$, define the quantity
$x(\gamma) := \frac{c(\gamma)}{\gamma}$. In view of
(\ref{eq:Equation1}) it is the unique solution to the equation
\begin{equation}
\frac{1}{\gamma} + x - (1+x) \log \left ( 1 + x \right )
 = 0,
\quad x > 0. \label{eq:Equation2}
\end{equation}
This equation is equivalent to
\begin{equation}
\frac{1}{\gamma} = \varphi(x), \quad x > 0 \label{eq:Equation3}
\end{equation}
where the mapping $\varphi : \mathbb{R}_+ \rightarrow
\mathbb{R}_+$ is given by
\[
\varphi(x) = (1+x) \log \left ( 1 + x \right ) - x , \quad x \geq
0.
\]
This mapping $\varphi : \mathbb{R}_+ \rightarrow \mathbb{R}_+$ is
strictly monotone increasing with $\lim_{x \downarrow 0}
\varphi(x) = 0$ and $\lim_{x \uparrow \infty} \varphi(x) =
\infty$, so that $\varphi$ is a bijection from $\mathbb{R}_+$ onto
itself. It then follows from (\ref{eq:Equation3}) that $x(\gamma)$
is strictly decreasing as $\gamma$ increases. Since $\varphi(1) =
\left ( \gamma^\star \right )^{-1}$, we get $x(\gamma^\star) = 1$
by uniqueness, whence $x(\gamma) < x(\gamma^\star) = 1$ for
$\gamma > \gamma ^\star$, a statement equivalent to
(\ref{eq:NeededInequality}).

Careful inspection of the proof shows that
(\ref{eq:KeyRingSizeBound}) holds with
\begin{equation}
h(\gamma;c) := - \max \left ( a(\gamma;c), b(\gamma;c) \right )
\label{eq:h}
\end{equation}
on the range $c(\gamma) < c < \gamma $, and it is clear from the
discussion above that $h(\gamma ; c ) > 0$ when $\gamma > \gamma
^\star$. \myendpf

\section*{Acknowledgment}
This work was supported by NSF Grant CCF-07290.

\bibliographystyle{IEEE}


\end{document}